\documentclass[twocolumn,aps,pra,superscriptaddress]{revtex4-1}
\usepackage[]{fontenc}
\usepackage{amsmath}
\usepackage{appendix}
\usepackage[inline]{enumitem}
\usepackage{multirow}
\usepackage{amssymb}
\usepackage{amsthm}
\usepackage{graphicx}
\usepackage{bbm}
\usepackage{color}
\usepackage{hyperref}
\definecolor{darkred}{rgb}{0,0,1}
\definecolor{darkgreen}{rgb}{0,0,1}
\definecolor{darkblue}{rgb}{0,0,1}

\hypersetup{colorlinks,breaklinks,linkcolor=darkred,citecolor=darkgreen,urlcolor=darkblue}

\def \diracspacing {0.7pt}
\newcommand{\ket}[1]{| \hspace{\diracspacing} #1 \rangle} 
\newcommand{\braket}[2]{\langle #1 \hspace{\diracspacing} | \hspace{\diracspacing} #2 \rangle} 
\newcommand{\norm}[2][]{#1\left| \! #1\left| #2 #1\right| \! #1\right|}

\newcommand{\abs}[2][]{#1| #2 #1|}

\newcommand{\unit}{\mathbb{I}}
\DeclareMathOperator{\tr}{tr}
\DeclareMathOperator{\sgn}{sgn}
\theoremstyle{definition}
\theoremstyle{plain}

\theoremstyle{remark}

\begin{document}
\title{Self-testing mutually unbiased bases in higher dimensions with space-division multiplexing optical fiber technology}
\author{M{\'a}t{\'e} Farkas}
\affiliation{Institute of Theoretical Physics and Astrophysics, National Quantum Information Centre, Faculty of Mathematics, Physics and Informatics, University of Gdansk, 80-952 Gdansk, Poland}
\affiliation{International Centre for Theory of Quantum Technologies, University of Gdansk, 80-308 Gdansk, Poland}
\author{Nayda Guerrero}
\affiliation{Departamento de F\'isica, Universidad de Concepci\'on, 160-C Concepci\'on, Chile}
\affiliation{Millennium Institute for Research in Optics, Universidad de Concepci\'on, 160-C Concepci\'on, Chile}
\author{Jaime Cari\~{n}e}
\affiliation{Departamento de Ingenier\'ia El\'ectrica, Universidad Cat\'olica de la Sant\'isima Concepci\'on, Concepci\'on, Chile}
\author{Gustavo Ca\~{n}as}
\affiliation{Departamento de F\'isica, Universidad del B\'io-B\'io, Collao 1202, Casilla 5C, Concepci\'on, Chile}
\author{Gustavo Lima}
\affiliation{Departamento de F\'isica, Universidad de Concepci\'on, 160-C Concepci\'on, Chile}
\affiliation{Millennium Institute for Research in Optics, Universidad de Concepci\'on, 160-C Concepci\'on, Chile}
%
\begin{abstract}
In the device-independent quantum information approach, the implementation of a given task can be self-tested solely from the recorded statistics and without detailed models for the employed devices. Even though experimentally demanding, it provides appealing verification schemes for advanced quantum technologies that naturally fulfil the associated requirements. In this work, we experimentally study whether self-testing protocols can be adopted to certify the proper functioning of new quantum devices built with modern space-division multiplexing optical fiber technology. Specifically, we consider the prepare-and-measure protocol of M.~Farkas and J.~Kaniewski (Phys.~Rev.~A 99, 032316) for self-testing measurements corresponding to mutually unbiased bases (MUBs) in a dimension $d>2$. In our scheme, the state preparation and measurement stages are implemented with a multi-arm interferometer built with new multi-core optical fibers and related components. Due to the high-overlap of the interferometer's optical modes achieved with this technology, we are able to reach the required visibilities for self-testing the implementation of two four-dimensional MUBs. We also quantify two operational quantities of the measurements: (i) the incompatibility robustness, connected to Bell violations, and (ii) the randomness extractable from the outcomes. Since MUBs lie at the core of several quantum information protocols, our results are of practical interest for future quantum works relying on space-division multiplexing optical fibers.
\end{abstract}

\maketitle

\section{Introduction}

The advent of quantum information technologies comes with promises such as exponential computational speed-up compared to currently existing classical algorithms \cite{Shor,Feynman}, or unconditionally secure quantum communication \cite{BB84,E91}. However, the success of these protocols relies on certification methods to verify that the used devices perform the tasks they are promised to. Furthermore, these verification methods should be possible to perform efficiently, and using only classical resources. Consequently, they are currently the subject of intensive study in the quantum information community \cite{ABGMPS07,GBHA10,BGLP11,RYZS12,HYN13,HBS13,Glima16,Glima19}, where they are commonly referred to as self-testing protocols. The strongest method is known as the ``device-independent approach'', which involves two parties sharing an entangled quantum state, and the only considered assumption for self-testing the proper implementation of a given task is that these parties are space-like separated. The certification is based solely on the recorded measurement statistics of the two parties \cite{self-test_review}. This method, however, comes with a few drawbacks. Firstly, it is rather challenging to implement experimentally, as it requires the production of entangled states with very high fidelities. Secondly, in dimensions larger than two, the theoretical treatment becomes complex as well. Accordingly, there are only a few theoretical results available for high-dimensional quantum states (qudits) \cite{selftest_qutrit,selftest_maxent}, and it has never been experimentally demonstrated.

Nonetheless, the use of qudit systems is advantageous for several quantum information tasks. For example, they allow for larger violations of Bell inequalities \cite{Kaszlikowski_2000}, improvement on quantum computation and communication complexity tasks \cite{Araujo_2014,Martinez_2018}, and higher randomness generation rates \cite{Glima20}. Thus, there is a current need of more practical self-testing protocols in higher dimensions. In order to alleviate the difficulties mentioned above, several relaxations of the demanding device-independent scheme have been introduced. One generic direction is to move to the experimentally less demanding prepare-and-measure scenario, in which instead of sharing an entangled state, one party prepares a state and sends it to the other party who then measures it. In this scenario, further reasonable assumptions are necessary to devise certification methods. These include bounds on the average energy of the quantum states \cite{SDIenergy}, or the indistinguishability of the different states prepared \cite{SDIoverlap}. Perhaps the most traditional such relaxation is to fix the dimension of the quantum states \cite{GBHA10,SDI2,SDI3,Glima14}.

Recently in Ref.~\cite{selftestMUB}, a method has been proposed for self-testing high-dimensional measurements corresponding to mutually unbiased bases (MUBs) in the prepare-and-measure scenario, under the dimension assumption. MUBs constitute a particularly useful family of quantum measurements, with myriad applications in quantum information. Among other tasks, they optimise state determination \cite{ivonovic,Glima11}, generate maximal amount of randomness \cite{maassen_uffink}, and give rise to secure cryptographic protocols \cite{BB84} (for a survey, see \cite{onMUBs}). The authors of Ref.~\cite{selftestMUB} anticipate that their certification method can be performed with currently available technologies in dimensions larger than two. This is precisely the aim of the current work, in which we experimentally study whether self-testing certification methods can be adopted in the new platform of space-division multiplexing (SDM) optical fiber technology to quantum information processing \cite{Glima20B}. In our scheme, we use single-photon path-encoded four-dimensional quantum systems (ququarts), and the state preparation and measurement stages are implemented resorting to an advanced four-arm interferometer built of multi-core optical fibers and new related technology \cite{Glima20}. As first observed in Ref.~\cite{Glima20}, this scheme should in principle attain the optical quality required for implementing self-testing protocols. Indeed, here in our test of Ref.~\cite{selftestMUB}, we are able to record the corresponding data with an average visibility of $99.89\%$, which allows us to self-test the proper implementation of a pair of four-dimensional MUBs. Moreover, we also certify the incompatibility of our implemented measurements, and the randomness extractable from their outcomes. Note that while this same type of protocol has already been implemented experimentally in higher dimensions \cite{QRACArmin,QRAC9,QRAC1024}, the error rates have never been suppressed to a level that would allow the self-testing of the measurements performed. Thus, our results clearly show the potential advantages of modern SDM technologies for high-dimensional quantum information processing.

\section{Theory}\label{sec:preliminaries}

Formally, a pair of MUBs in dimension $d$ corresponds to two rank-1 measurements projecting onto the orthonormal bases $\{\ket{a_i}\}_{i = 1}^d$ and $\{\ket{b_j}\}_{j = 1}^d$ on $\mathbb{C}^d$. We say that these bases are mutually unbiased if
\begin{equation}\label{eq:MUB}
\abs{\braket{a_i}{b_j}}^2 = \frac1d~~\forall i,j\in\{1,\ldots,d\},
\end{equation}
that is, every pair of vectors from different bases has the same overlap. One simple example is the eigenbases of the Pauli $X$- and $Z$-operators on a qubit.

The prepare-and-measure self-testing method used in Ref.~\cite{selftestMUB} to certify $d$-dimensional MUB measurements is based on the so-called $2^d\to1$ quantum random access code (QRAC) protocol. In a QRAC, the preparation side (Alice) receives two uniformly random classical dits, $i,j\in\{1,\ldots,d\}$. Based on this input, Alice prepares the $d$-dimensional quantum state, $\rho_{ij}$, and sends it to Bob on the measurement side. Bob receives a uniformly random classical bit, $y\in\{1,2\}$, based on which he decides which observable to measure on the state $\rho_{ij}$. If $y=1$, his measurement is a $d$-outcome positive operator-valued measure (POVM), whose measurement operators are denoted by $A_i$. Similarly, for $y=2$, he measures $B_j$. Recall that for POVMs we have that $A_i,B_j\ge0$ and $\sum_{i = 1}^dA_i = \sum_{j = 1}^d B_j =\unit$. That is, a $d$-outcome POVM is a set of $d$ positive semidefinite operators that add up to the identity operator. Let us denote the outcome of Bob's measurement by $b\in\{1,\ldots,d\}$. The parties' common aim is that when $y=1$, Bob's output equals Alice's first input, that is, $b=i$, and when $y=2$ they have $b=j$. To quantify their success rate, we employ the \emph{average success probability} (ASP)
$\bar{p}=\frac12\big[P(b=i|y=1)+P(b=j|y=2)\big]$ as the figure of merit. According to the Born rule, the probability of Bob outputting $b$ when Alice's input is $i,j$ and Bob's input is $y = 1$ is $\tr( \rho_{ij} A_b )$, and similarly it is $\tr( \rho_{ij} B_b )$ when $y = 2$. That is, the ASP for a generic encoding scheme $\rho_{ij}$ and measurement choice $A_i$ and $B_j$ can be written as
\begin{equation}\label{eq:asp2}
\bar{p} = \frac{1}{2d^2} \sum_{i,j = 1}^d \tr \big[ \rho_{ij} (A_i + B_j) \big].
\end{equation}

In Ref.~\cite{selftestMUB} the authors provide certificates for MUBs based only on the recorded ASP in a QRAC. They show that in dimension $d$, \mbox{$\bar{p}\le\frac12\left(1+ \frac{1}{\sqrt{d}}\right)=:\bar{p}_Q$,} and this maximum can only be attained if Bob's measurements correspond to a pair of MUBs. Moreover, even for suboptimal $\bar{p}$ one can certify the closeness of the employed measurements to a pair of MUBs. Specifically, one can bound the entropy of the generalized overlaps of the two measurements and the sum of the individual operator norms. These two measures together -- having sufficiently high values -- imply that the measurement operators have close to uniform overlaps and are close to being rank-1 projectors, that is, they are close to MUBs.

Specifically, the first quantity employed is the \emph{overlap entropy}, $H_S(A,B) = H_\frac12\left(\left\{\tr(A_iB_j)/d\right\}_{ij}\right)$, where $H_\frac12$ is the $\frac12$-R\'enyi entropy (note that for projective measurements $\tr(A_iB_j) = \abs{\braket{a_i}{b_j}}^2$). It has been shown that given an observed ASP $\bar{p}$, it holds for the measurements $A$ and $B$ that \cite{selftestMUB}
\begin{equation}\label{eq:HSbound}
H_S(A,B) \ge 2 \log\left[d\sqrt{d}(2\bar{p}-1)\right].
\end{equation}
The maximal possible value of the overlap entropy, $\log d^2$, is reached by MUBs, and can be certified upon observing $\bar{p}=\bar{p}_Q$.

The second quantity is the \emph{sum of the norms}, \mbox{$N(A) = \sum_{i = 1}^d\norm{A_i}$}. It has also been shown in Ref.~\cite{selftestMUB} that
\begin{equation}\label{eq:Nbound}
N(A) \ge d-\frac{2+\sqrt{2}}{d}\left(1-\sqrt{d^3(2\bar{p}-1)^2-(d^2-1)}\right),
\end{equation}
and the same holds for $B$. The maximal possible value of the sum of the norms, $d$, is reached if and only if the measurements are rank-1 projective, and this can be certified upon observing $\bar{p}= \bar{p}_Q$.

Putting the above two bounds together, observing \mbox{$\bar{p} = \bar{p}_Q$} implies that $\tr(A_iB_j) = \frac1d$ for all $i,j$ and that the measurements are rank-1 projective. In other words, $\bar{p} = \bar{p}_Q$ certifies that the measurements of Bob correspond to a pair of MUBs. By the continuity of the bounds in $\bar{p}$, it follows that if the observed ASP is suboptimal, $\bar{p} < \bar{p}_Q$, but close to optimal, then the overlap entropy and the sum of the norms are both close to their unique MUB values. This serves as a certificate that the employed measurements are close to MUBs.

Last, the authors of Ref.~\cite{selftestMUB} derive certificates for two useful properties of the measurements: incompatibility robustness, and the amount of randomness generated. The former, briefly speaking, is the maximal visibility of the measurements at which they are jointly measurable (compatible) \cite{incomp}. Clearly, for compatible measurements pairs, this maximal visibility is 1, and the lower the value, the more incompatible the pair is. Jointly measurable observables are of no use in nonlocal and steering scenarios \cite{QVB14}, and therefore it is important to quantify the extent to which a pair of measurements is incompatible. In Ref.~\cite{selftestMUB}, the authors show that the incompatibility robustness of $A$ and $B$ is bounded by
\begin{equation}\label{eq:etabound}
\eta^\ast \le \frac{\frac12 d^2(1+s_\text{max})-\frac{N(A)^2}{d}}{N(A)^2-d-[d-
N(A)][d-N(A)+1]},
\end{equation} where $s_\text{max}=\max_{ij}\norm{\sqrt{ A_i}\sqrt{B_j}}$. Using the bounds in Eqs.~\eqref{eq:HSbound} and \eqref{eq:Nbound}, one can then bound the incompatibility robustness by the observed ASP. The value corresponding to a pair of MUBs, \mbox{$\eta^\ast = \frac12\left(1+\frac{1} {\sqrt{d}+1}\right)$} can be certified upon observing $\bar{p}= \bar{p}_Q$.

The second quantity to certify is the amount of uncertainty produced in the outcome of the measurements, formulated as an entropic uncertainty relation \cite{CBTW17}. This amounts to a lower bound on the entropy of the measurement outcome probabilities in a state-independent fashion. Let us denote the Shannon entropy of the outcome probabilities of the measurement $A$ on the state $\rho$ by $H(A)_\rho$. Then, it was shown in Ref.~\cite{selftestMUB} that given a QRAC ASP $\bar{p}$, it holds that
\begin{equation}\label{eq:HABbound}
\begin{split}
& \left. H(A)_\rho + H(B)_\rho \ge \right. \\
& \left.  -2\log \left(2\bar{p}-1+\frac1d\sqrt{d(d^2-1)[1-d(2
\bar{p}-1)^2]}\right) \right. ,
\end{split}
\end{equation} for any state $\rho$. Note that the maximal value for \mbox{rank-1} projective measurements, $\log d$, can again be certified upon observing $\bar{p}= \bar{p}_Q$.

\section{Space-division multiplexing technology}\label{sec:SDM}

\begin{figure}[h]
\centering
\includegraphics[width=.5\textwidth]{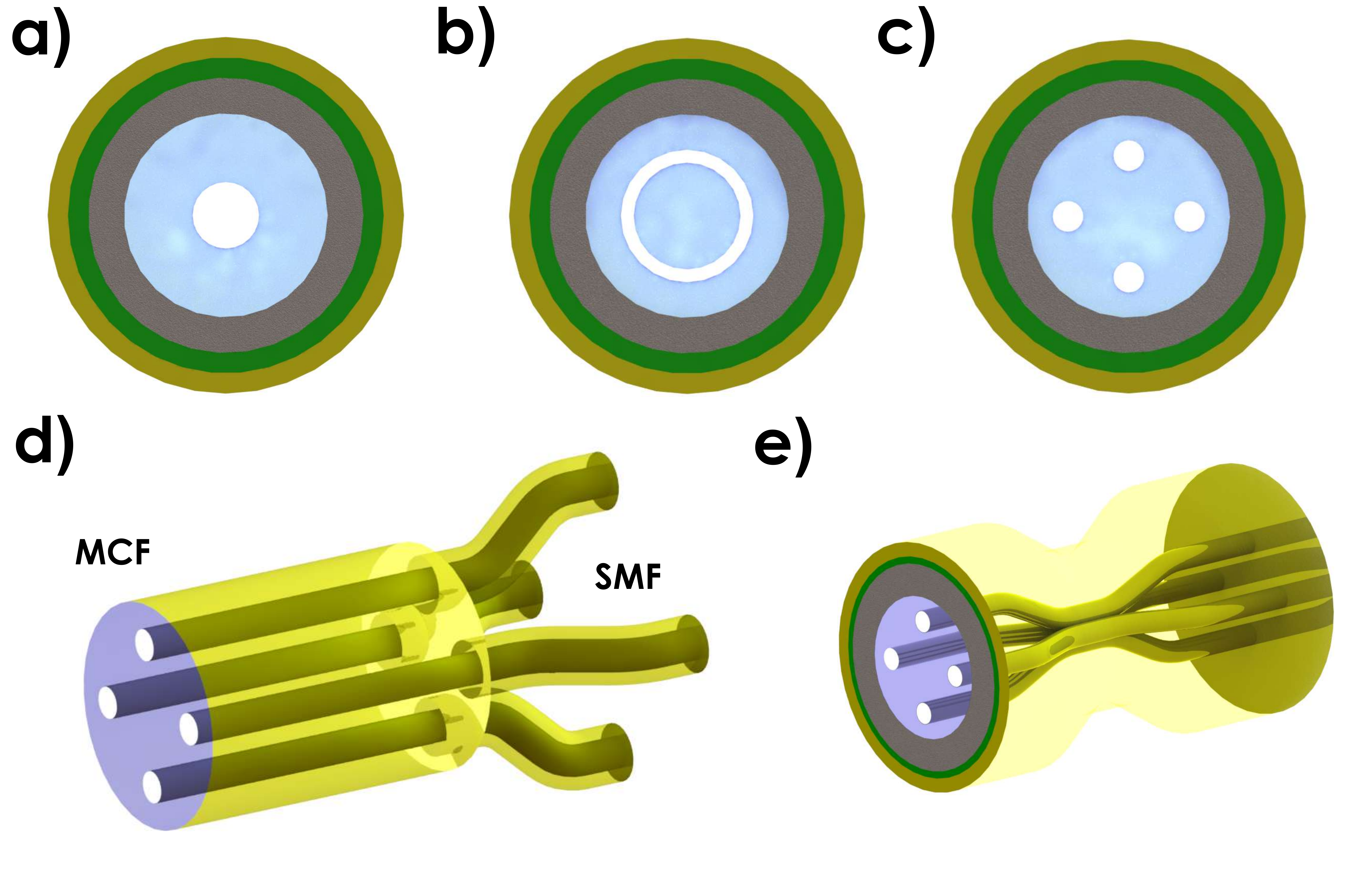}
\caption{Fibers and components typically used in the new SDM approach for fiber-optics communication. Schematic representation of the cross-section of a) Few-mode fiber, b) Ring-core fiber, and c) Multi-core fiber (with 4 cores). Each of them is composed of a core (white), cladding (light-blue), coating (grey), strength member (green), and outer jacket (yellow). FMF supports the propagation of a few linearly polarized modes. RCF has an annular core that supports, for instance, the propagation of some Laguerre-Gaussian optical modes. MCF is a single fiber with several single-mode cores within its cladding. d) Schematics of demultiplexer devices used for efficiently coupling light into multi-core fibers (insertion loss $<0.7$ dB). e) An example of a 4-core fiber integrated multi-port beamsplitter.} \label{Fig_SDM}
\end{figure}

Space-division multiplexing is a classical telecommunication technique that uses multiple transverse optical modes for increasing data communication capacity. The SDM technique is implemented for optical communication links in both free space and fiber optics \cite{OSDM,Richard13}. It is considered a crucial solution to overcome the so-called ``capacity crunch'' of fiber-optic communications \cite{Richard13}. In this case, SDM technology is typically based on few-mode fibers (FMFs) \cite{Sillard,Rademacher}, ring core fibers (RCFs) \cite{RCF}, and multi-core fibers (MCFs) \cite{Inao,Saitoh}. They are schematically represented on Fig.~\ref{Fig_SDM}.

FMFs are a particular class of multi-mode fibers (MMFs), which support only a few linearly polarized transverse optical modes \cite{Sillard,Rademacher}. Each mode that is supported in a FMF have very low cross talk to the others and, therefore, can be used as an independent data channel. In contrast, in a MMF all modes are combined together due to decoherence induced mode coupling effects, such that only one data channel can be accessed \cite{FMF_Ken}. RCFs are optical fibers with an annular refractive index profile that supports multiple Laguerre-Gaussian beams carrying orbital angular momentum (OAM) \cite{RCF_Wang}. Last, there are MCFs, that are considered to be the most promising solution for SDM, since their fabrication is cost effective \cite{Saitoh}. An MCF is a single fiber containing multiple cores within the same cladding, which are sufficiently separated from each other to avoid light coupling between them. Typically, cross-talk between the cores is negligible with more than 60dB of attenuation \cite{Saitoh}.

Together with the development of these fibers, there have been several new related components built for improving the efficiency of SDM techniques. For instance, there are multiplexer/demultiplexer (DEMUX) devices used to combine and separate the different transverse optical modes supported by the SDM fiber. Typically, these devices have $N$ independent single-mode fibers connected to the SDM fiber, mapping $N$ transverse Gaussian modes onto the $N$ particular optical modes supported by the SDM fiber. For FMFs, DEMUXs called photonic lanterns are used \cite{FMF_Demux}. For RCFs these devices are called mode sorters. They are usually built with bulk optics \cite{RCF_DemuxFree}, but an important recent development is an all-fiber mode sorter \cite{RCF_DemuxFiber}. Finally, the DEMUXs used for MCFs are devices composed of single-mode fibers (SMFs), each one connected to one core of the MCF. These devices are already commercially available and are built using a fiber-bundle polishing and tapering technique presented in \cite{MCF_Demux1,MCF_Demux2}. In Fig.~\ref{Fig_SDM}(d) we show, as an example, the schematics of a MCF DEMUX.

Our experimental setup is based on MCFs and in this case another important device is the multi-core fiber integrated multi-port beam-splitter (MBS), recently presented in Ref. \cite{Glima20} (See Fig.~\ref{Fig_SDM}(e)). This device is crucial for quantum information processing  because it allows one to implement  distinct unitary operations representing the change of basis from the logical basis to a basis mutually unbiased to it. Thus, it allows for the generation and measurement of a general class of quantum states, as we explain in the next section. The MBS is fabricated directly within a multi-core fiber using a novel tapering technique for MCFs \cite{Ming}. By tapering the fiber, the cores are brought together, and due to evanescent effects, there is light coupling from one core to the others. Due to the symmetry of the MCF structure, the splitting ratio can be made balanced for all core-to-core combinations.

\section{Experiment}\label{sec:experiment}

Recently, the new technology developed for SDM has become a new platform for high-dimensional quantum information processing \cite{Glima20B}. Initial efforts, based on path-encoded qudits and multi-core fibers \cite{Shields,Glima17,Bacco17,Lee17,Esteban20}, have now been expanded to different types of fibers and encoding schemes \cite{Cui17,Sit18,Cao18,Cozzolino19}. Nonetheless, up to our knowledge, this new platform has not yet been demonstrated to be compatible with modern self-testing protocols of quantum states and circuits. Here, we fill this gap by extensively studying the protocol of Ref.~\cite{selftestMUB}. Specifically, we measure the QRAC ASP, and bound all the quantities of Eqs.~\eqref{eq:HSbound}--\eqref{eq:HABbound} with a four-arm Mach--Zehnder (MZ) interferometer built of MCFs and related technology.

\begin{figure}[t]
\centering
\includegraphics[width=1.1\linewidth]{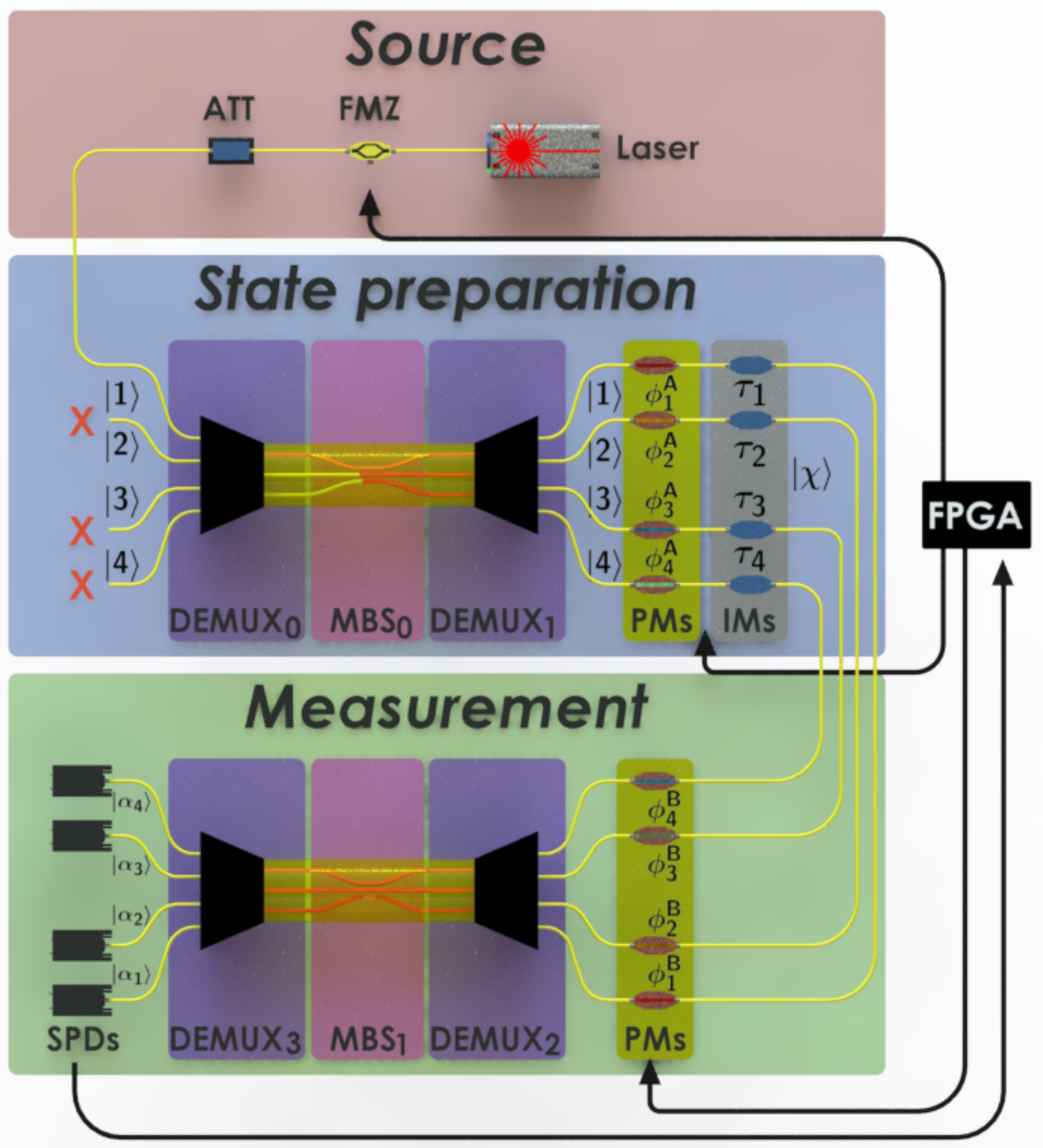}
\caption{The experimental setup is based on a four-arm Mach-Zehnder interferometer built of four-core multi-core fibers (MCF) and related technology. The interferometer is used for preparing and measuring path-encoded ququart states. At the state preparation stage, the initial state is prepared by a set composed of a $4\times4$ MCF based multi-port beam splitter (MBS$_0$), Phase (PM), and amplitude (IM) fiber-pigtailed modulators. The measurement is achieved using another set of PMs and a second MBS$_1$ connected to four InGaAs single-photon detection (SPD) modules. The field-programmable gate array (FPGA) electronic unit automatically controls the protocol implementation. See the main text for details.} \label{Fig_setup}
\end{figure}

The state preparations in the QRAC protocol are realized by photonic states.
The initial photon source is a continuous-wave telecom laser, operating at 1546 nm (see Fig.~\ref{Fig_setup}). It is connected to an external fiber-pigtailed amplitude modulator (FMZ), which is controlled by a field-programmable gate array electronic unit (FPGA) to generate 5 ns wide pulses. Then, we use optical attenuators (ATT) to create weak coherent states. The attenuators are calibrated to set the average number of photons per pulse to $\mu=0.2$. In this case, the probability of having pulses containing at least one photon is $P(\mu=0.2| n\geq 1)\approx 18\%$. Most of the non-null pulses contain only one photon, and represent $90.3\%$ of the experimental runs. Therefore, our source can be seen as a good approximation of a non-deterministic source of single photons \cite{GisinQC2002}.

The source's signal is sent to a commercial fiber built-in DEMUX unit (DEMUX$_0$), which consists of four independent single-mode fibers, each of them connected to one core of a four-core MCF. The source is connected through one of the four SMFs of DEMUX$_0$, therefore only one core of the MCF is illuminated. DEMUX$_0$ is then connected to a MCF-based $4\times4$ MBS (denoted MBS$_0$), whose matrix representation is given by \cite{Glima20}
\begin{equation}\label{Ec_ES_01}
\text{MBS}=
\frac{1}{2} \begin{bmatrix}
     1&1&1&1\\
		1&1&-1&-1\\
		1&-1&1&-1\\
		1&-1&-1&1
\end{bmatrix}
\end{equation} in the logical basis. In our scheme, the logical states are defined in terms of the core modes available for the photon propagation over the multi-core fiber \cite{Glima17,Glima20}. Therefore, the $4\times4$ MBS corresponds to a Hadamard gate in dimension four.

MBS$_0$ is then coupled to a second DEMUX (denoted DEMUX$_1$) via their respective MCFs. In order to control the initial quantum state entering the interferometer, we connect phase (PM) and amplitude (IM) fiber-pigtailed modulators to each SMF of DEMUX$_1$, which are controlled by the FPGA. The general path-encoded ququart state that we can prepare in the first part of the MZ is then given by
\begin{equation}\label{Ec_ES_03}
 | \chi  \rangle = \frac{1}{\sqrt{N}}\sum^{ 4}_{k=1} \tau_k e^{i \phi^A_k}  | k  \rangle,
\end{equation} where $\ket{k}$ represents the state of the photon transmitted in the $k$-th core (i.e.~the $k$-th logical state). $\tau_k$ and $\phi^{A}_k$ are the transmissivity and relative phase, respectively, of core $k$, and $N$ is the normalization constant.

Having prepared the state, the measurements are performed in a similar fashion, using a second set of PMs, DEMUX$_2$ and MBS$_1$ \cite{Glima20}. The resulting unitary operation implemented is
\begin{equation}\label{Ec_ES_04}
U_M=\frac{1}{2} \begin{bmatrix}
 e^{i\phi_{1}^{B}} & e^{i\phi_{1}^{B}} & e^{i\phi_{1}^{B}} & e^{i\phi_{1}^{B}} \\
 e^{i\phi_{2}^{B}} & e^{i\phi_{2}^{B}} & -e^{i\phi_{2}^{B}} & -e^{i\phi_{2}^{B}} \\
 e^{i\phi_{3}^{B}} & -e^{i\phi_{3}^{B}} & e^{i\phi_{3}^{B}} & -e^{i\phi_{3}^{B}} \\
 e^{i\phi_{4}^{B}} & -e^{i\phi_{4}^{B}} &  -e^{i\phi_{4}^{B}} & e^{i\phi_{4}^{B}} \\\end{bmatrix},
\end{equation} where $\phi^{B}_{k}$ is the phase applied in the core mode $k$ at the measurement side. After applying the phases, we conclude the projective measurement using a final DEMUX (denoted DEMUX$_3$), to send the outcomes of MBS$_1$ to four single-photon detectors (SPD). The detectors are triggered commercial InGaAs single-photon detection modules, configured with 5 ns detection gate and $10 \%$ of detection efficiency. The detection counts are recorded by the FPGA unit. The measurement corresponding to the above procedure is the rank-1 projective measurement given by the states
\begin{equation}\label{eq:expbasis}
\begin{split}
|\alpha_{1}\rangle & \left. = \frac{1}{2}(e^{i\phi_{1}^{B}} |1 \rangle+e^{i\phi_{2}^{B}} |2 \rangle+e^{i\phi_{3}^{B}} |3 \rangle+e^{i\phi_{4}^{B}} |4 \rangle) , \right. \\
|\alpha_{2}\rangle & \left. = \frac{1}{2}(e^{i\phi_{1}^{B}} |1 \rangle+  e^{i\phi_{2}^{B}} |2 \rangle- e^{i\phi_{3}^{B}} |3 \rangle- e^{i\phi_{4}^{B}} |4 \rangle), \right. \\
|\alpha_{3}\rangle & \left. = \frac{1}{2}(e^{i\phi_{1}^{B}} |1 \rangle- e^{i\phi_{2}^{B}} |2 \rangle + e^{i\phi_{3}^{B}} |3 \rangle - e^{i\phi_{4}^{B}} |4 \rangle), \right. \\
|\alpha_{4}\rangle & \left. = \frac{1}{2}(e^{i\phi_{1}^{B}} |1 \rangle-  e^{i\phi_{2}^{B}} |2 \rangle - e^{i\phi_{3}^{B}} |3 \rangle + e^{i\phi_{4}^{B}} |4 \rangle). \right.
\end{split}
\end{equation}
That is, photon detection in path $k$ corresponds to the measurement outcome associated with $| \alpha_k \rangle$.

Therefore, in our experiment we can prepare the state of Eq.~\eqref{Ec_ES_03} and measure it in the basis defined by the orthorgonal states of Eq.~\eqref{eq:expbasis}. Fiber-based polarization controllers (not shown for the sake of simplicity) are used in each path to guarantee the indistinguishability of the core modes, such that there is no path-information available to compromise the visibility of the interferometer \cite{QErasure,DecQErasure}.

The FPGA unit controls and synchronises the preparation and measurement stages, with both working at a repetition rate of 2 MHz. Due to thermal and mechanical fluctuations in the interferometer, time-dependent phase noise occurs. To overcome this, we model the phase applied at the preparation stage in each arm as $ \phi^A_k = \phi^n_k + \phi^c_k + \phi^s_k $, where $ \phi^n_k $ represents the phase noise, $\phi^c_k $ the phase of a noise suppressor that we control by a continuous low-speed voltage signal, and $ \phi^s_k $ the phase used to prepare the desired state, which we control by a high-speed voltage. Both voltages are controlled by the FPGA unit through a power driver. To cancel the phase noise, a control algorithm in the FPGA sets $\tau_k=1$ and the high-speed voltage to 0, while perturbing $\phi^c_k$ to maximize the single-counts at SPD$_1$. Thus, preparing the state  $| \chi_c  \rangle =\frac{1}{2 }\left( | 1  \rangle+   | 2  \rangle+  | 3  \rangle+ | 4  \rangle\right)$. Once a given threshold of single-counts is reached, $ \phi^c_k $ is kept fixed, and the experiment is performed using the fast-switching phases $ \phi^s_k $ for preparing the states, and $\phi^B_k$ for preparing the measurements considered in the protocol. Typically, 60000 detections are recorded over 1 s of integration time.

The measurements, which we aim to certify, correspond to a pair of MUBs, which we chose such that they can be implemented in our setup using only phase modulation, without the need of amplitude modulation. Specifically, the two bases $\{\ket{a_i}\}_{i = 1}^4$ and $\{\ket{b_j}\}_{j = 1}^4$ are given by the columns of the matrices
\begin{eqnarray}
A&=&\frac{1}{2}\begin{bmatrix}
1 & 1 & 1 & 1\\
1 & 1 & -1 & -1 \\
1 & -1 & 1 & -1 \\
1 & -1 & -1 & 1
\end{bmatrix}, \\
B&=&\frac{1}{2}
\begin{bmatrix}
-1 & -1 & -1 & -1 \\
1 & 1 & -1 & -1 \\
1 & -1 & 1 & -1 \\
1 & -1 & -1 & 1
\end{bmatrix},
\end{eqnarray}

According to Eq.~\eqref{Ec_ES_04}, $A$ can be performed by setting the phases $ \phi_{k}^{B} = 0 $ for all $ k = 1,2,3,4 $, while $B$ can be performed by setting $ \phi_ {1}^B = \pi $ and keeping the other phases equal to zero. In the QRAC protocol described above, Bob chooses the measurement basis $A$ or $B$ according to his input $ y $. In our experiment, we perform this basis choice simply by changing $ \phi_{1}^{B} $: when $ y = 1 $, we choose $ \phi_{1}^{B} =  0 $, and when  $ y =   2  $, we choose $  \phi_{1}^B = \pi $.

\begin{figure*}[t]
\includegraphics[width=0.85\linewidth]{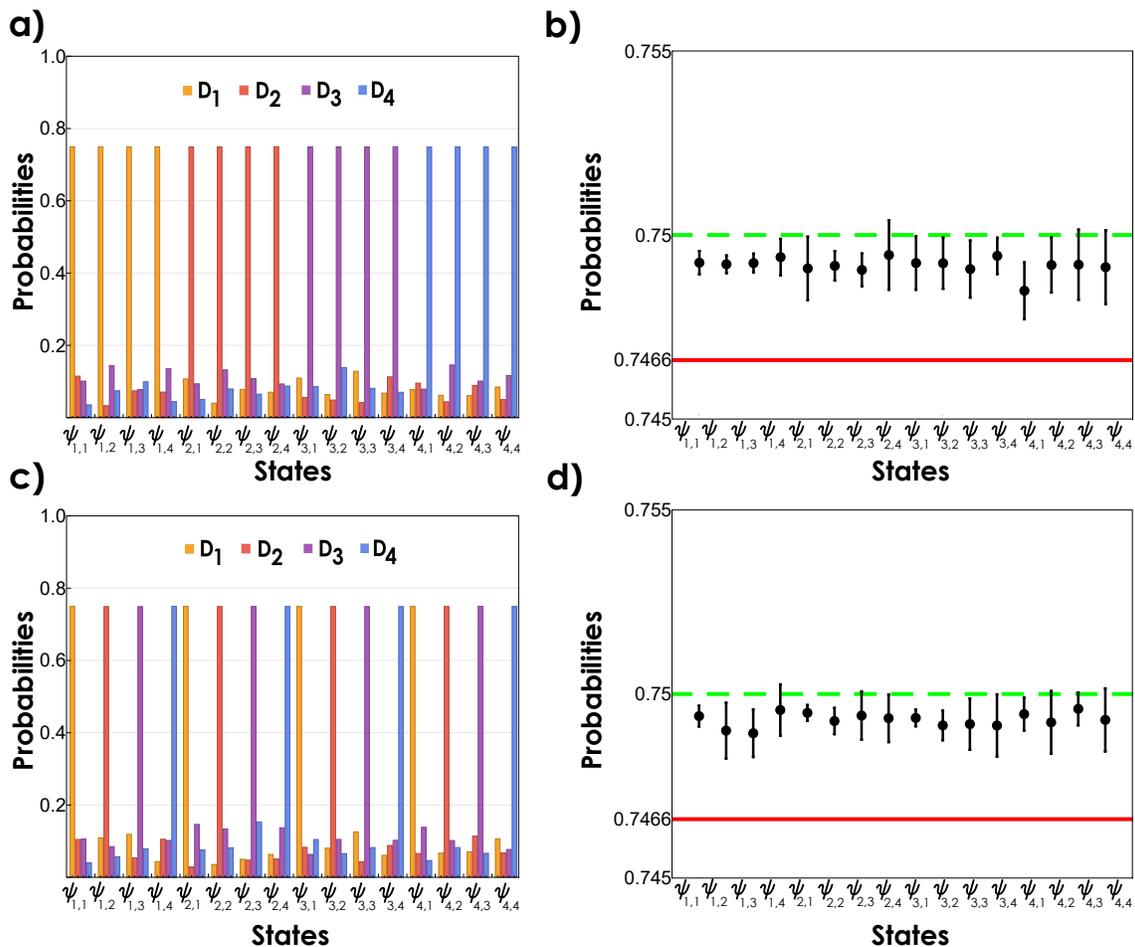}
\caption{In a) and c) we show the outcome probabilities of the measurements $A$ and $B$, respectively, for each state $|\psi_{ij}\rangle$. In b) and d) we show the ASP for each state $|\psi_{ij}\rangle$ upon measuring $A$ and $B$, respectively. The green line corresponds to the optimal ASP $\bar{p} = 0.75$, while the red line corresponds to the minimal ASP such that the most demanding quantity, $\eta^\ast$, can be self-tested.}
\label{MUB_A_B}
\end{figure*}

The optimal state preparation for Alice's input $i,j$ is the pure state \cite{selftestMUB}:
\begin{equation}\label{Prob05}
\left  |  \psi_{ij} \right \rangle=\frac{1}{ \sqrt{3}}
(\left  | a_i \right \rangle+ \sgn\left(   \left\langle   a_i | b_j \right \rangle  \right) \left  | b_i \right \rangle),
\end{equation}
which we can produce according to Eq.~\eqref{Ec_ES_03}. The QRAC protocol is then carried out by randomly preparing the 16 different states $\ket{\psi_{ij}}$ with $i,j \in \{1,2,3,4\}$, randomly measuring them in the bases $A$ or $B$, and collecting the measurement statistics to estimate the average success probability in \eqref{eq:asp2}. The choice of states and measurements are implemented directly into the FPGA resorting to a pseudo-random number generation algorithm.

\section{Results}\label{sec:results}

We present the recorded experimental data in two parts, corresponding to the success probabilities related to the measurement $A$ and $B$ of Eq.~\eqref{eq:asp2}. Fig.~\ref{MUB_A_B} a) contains the outcome probabilities for the interferometer's outcomes 1, 2, 3 and 4 for each state $\ket{\psi_{ij}}$ upon measuring $A$. In Fig.~\ref{MUB_A_B} b) we show the ASP for each state $|\psi_{ij}\rangle$ upon measuring $A$. On average, we observe an ASP of $\bar{p}_A=0.7491 \pm 0.00017$ for this measurement, where the error is calculated using the Poissonian distribution for the number of photon detections. The analogous data for the measurement $B$ is depicted on Figs.~\ref{MUB_A_B}~c)~and~d), yielding an ASP of $\bar{p}_B=0.7493 \pm 0.0001$ in this case.

Putting the above values together, the overall ASP is \mbox{$\bar{p} = 0.74924 \pm 0.00011$.} Using this result and Gaussian error propagation, from \eqref{eq:HSbound} we obtain that \mbox{$H_S(A,B) \ge 3.99122 \pm 0.00131$}. From \eqref{eq:Nbound} we get that \mbox{$N(A) \ge 3.95749 \pm 0.00649$}. These two results, together, self-test that the measurements are close to a pair of MUBs ($H_S(A,B) = 4$ and $N(A) = 4$).

Concerning the operational quantities, from \eqref{eq:etabound} we obtain
\mbox{$\eta^\ast \le 0.798757 \pm 0.010997$.} Therefore, we certify a non-trivial
bound on the critical visibility of our measurements at which they become
compatible. This confirms that the measurements used in the experiment are
indeed incompatible and therefore will be useful in future Bell and steering experiments \cite{Esteban20}.

Lastly, from \eqref{eq:HABbound} we can bound the entropic uncertainty: \mbox{$H(A)_\rho + H(B)_\rho \ge
1.24581 \pm 0.04886$.} That is, we obtain a minimal entropy that can be extracted from the
outcomes of our measurements on \emph{any} quantum state. This can be used for
secure random number generation or quantum key distribution protocols.

\section{Conclusions}\label{sec:conclusions}

With the development of new quantum technologies, there is a current need of certification schemes for preparing high-dimensional quantum states and measurements. Since self-testing methods in nonlocal scenarios are complicated both in theory and practice, recently proposed methods for self-testing quantum devices in the prepare-and-measure scenario become relevant. In this work, we demonstrate the viability of adopting such type of protocols in higher-dimensions to validate the proper functioning of new quantum devices built with modern space-division multiplexing technology. Specifically, we self-test the proper implementation of measurements corresponding to mutually unbiased bases in dimension $d=4$.

Our results show that space-division multiplexing is an advantageous platform for high dimensional quantum information processing, achieving an exceptionally high optical quality with visibilities greater than $99\%$. While experiments implementing the same protocol have previously been performed \cite{QRACArmin,QRAC9,QRAC1024}, our technique allowed us not only to certify the quantum advantage in random access coding, but to self-test the measurements under the dimension assumption, as well as to certify their level of incompatibility and the amount of randomness extractable from their outcomes. These results are of practical relevance for future experiments relying on this technology, since mutually unbiased measurements lie at the core of several quantum information protocols.

\section*{Acknowledgements}
This work was supported by Fondo Nacional de Desarrollo Científico y Tecnológico (Conicyt) (Fondecyt 3170596, 1190933, and 1200859), and Millennium Institute for Research in Optics. MF acknowledges support from the Polish NCN grant Sonata~UMO-2014/14/E/ST2/00020 and from the project of the Polish National Agency for Academic Exchange “International scholarship exchange of PhD candidates and academic staff”, project No. POWR.03.03.00-IP.08-00-P13/18. 

\bibliography{expbib}

\bibliographystyle{ieeetr}




\end{document}